# Tunable Coatings on Various Substrates for Self-Adaptive Energy Harvesting with Daytime Solar Heating and Nighttime Radiative Cooling

*Ken Araki, Vishwa Krishna Rajan, and Liping Wang\**

K. Araki, V. K. Rajan, L. Wang
School for Engineering of Matter, Transport and Energy, Arizona State University, 501 E Tyler Mall, Tempe, AZ, 85287, USA
E-mail: liping.wang@asu.edu

Funding: NSF Grant No. CBET-2212342

**Abstract**: In this work, tunable vanadium dioxide ($VO_2$) metafilms on different substrate materials fabricated via low-oxygen furnace oxidation are demonstrated for self-adaptive daytime solar heating and nighttime radiative cooling. Because of its thermally-driven insulator-to-metal phase transition behavior, the $VO_2$ metafilms work as spectrally-selective solar absorber with a high solar absorptance of 0.86 and a low infrared emissivity of ~0.2 at daytime, while they behave as selective cooler at nighttime to dissipate heat effectively through the atmospheric transparency window with a high emissivity of ~0.76 to cold outer space. From the outdoor vacuum tests, a significant temperature rise up to 169 K upon solar heating and a temperature drop of 17 K at night are experimentally observed from these tunable $VO_2$ metafilms. With the atmosphere temperature fitted in-situ, the accurate heat transfer model shows excellent agreement with the stagnation temperature measurement, and indicates a high heating power of ~400 W/m$^2$ at 80°C sample temperature in the middle of the day, and a cooling power of ~60 W/m$^2$ at 30°C in equilibrium with ambient at night. This work would facilitate the development of self-adaptive coatings with cost-effective and scalable fabrication approaches for all-day energy harvesting.

# 1. Introduction

Sustainable and renewable energy sources are in pressing needs due to escalating demand for energy worldwide and ever-present concern of global warming.[1] Net-zero energy consumption is recognized as an ultimate goal using nanoscale materials and radiative heat transfer technologies.[2] Solar energy ($\lambda$ = 0.3~2.5 μm) and thermal radiation (2.5~100 μm) are both useful resources that can extract electrical power using various technologies such as solar photovoltaic,[3] solar-thermoelectric,[4] radiative-thermoelectric,[5] and thermoradiative[6] power generations. Regardless of decades of development in such devices for improved energy efficiency and power output,[7] it is still a daunting challenge to continuously produce power with these energy sources all day long.

Continuous power output using thermoelectric generator (TEG) with broadband black absorber/emitter was proposed but the temperature difference across TEG was much limited due to excessive radiative loss during daytime and high radiative absorption from atmosphere at night.[5] To maximize the solar thermal energy harvesting at daytime, spectrally selective solar absorber is required, which has been extensively researched over last decade in various novel photonic crystal, metafilm, metamaterial and compound structures.[8] To maximize energy dissipation at night, selective coolers that only emit within atmospheric transparency window (8 ~ 14 μm) is sought after, for which an average of −37 K temperature drop has been experimentally demonstrated in a vacuum environment under certain weather conditions.[9] Another experimental study demonstrated simultaneous solar heating with 24°C above the ambient and radiative cooling with −29°C temperature drop below ambient by placing an infrared transparent solar absorber on top of ZnSe viewport above the radiative cooler inside the vacuum.[10]

Tunable coatings, which would behave as selective solar absorbers at daytime and selective emitter at night, could boost the continuous energy-harvesting performance by taking advantage of both solar thermal and radiative cooling effects. The tunability in optical and radiative properties can be realized with external stimuli such as mechanical stress [11] and electrical means,[12] or preferably self-adaptively with temperature. Thermochromic materials,[13] in particular $VO_2$,[14] whose insulator-to-metal (IMT) phase transition occurs around 68°C, show great promise for this purpose. While it has been widely studied for smart windows[15-19], infrared camouflage,[20] switchable infrared filters,[21, 22] thermal diode[23-25] and radiative thermal control applications,[26-31] very few have demonstrated self-adaptive $VO_2$ coatings with daytime solar heating and nighttime radiative cooling. Ao et al. fabricated $VO_2$ thin film on lattice-matching sapphire substrate via molecular beam epitaxy (MBE), which exhibits high solar absorptivity of 0.89 and emissivity contrast of 0.50 within the atmospheric transparency window.[32] Temperature differences of 170 K and -20 K to the ambient temperature ($T_{amb}$ ~ 15°C at daytime and ~8°C at nighttime) were respectively demonstrated at daytime and nighttime with outdoor tests in vacuum. However, different approaches need to be adopted to fabricate $VO_2$ on different substrate materials. Liu et al. used wet oxidation method to synthesize $VO_2$ on quartz substrate, where they demonstrated 95 K temperature rise at daytime and 5 K temperature drop with cooling power of 58 W/m$^2$ in ambient outdoor test.[33] Rensberg et al. fabricated $VO_2$ thin films on transparent conductive oxides (TCOs) using magnetron sputtering, which was reflective in the infrared with metallic $VO_2$ but emissive from TCOs at with transparent insulating $VO_2$. As a result, spectral infrared reflectivity change of 0.7 at 10 μm wavelength was achieved.[34] Similarly, $VO_2$ thin films were also deposited on flexible polymer substrate via reactive magnetron sputtering, whose infrared emissivity changes 0.32 to 0.82.[35]

Growth of high-quality $VO_2$ thin films is not trivial and it is more difficult to fabricate tunable self-adaptive energy harvesting coatings with dramatic change in optical properties by integrating $VO_2$ with other materials on arbitrary substrate. In this work, we demonstrate cost-effective fabrication of tunable $VO_2$ metafilm coatings on various substrate materials, i.e., Quartz, undoped-silicon (UDSi), and aluminum (Al), and superior performance in daytime solar thermal energy harvesting and nighttime radiative cooling from outdoor vacuum tests. In particular, high-quality $VO_2$ thin film is thermally grown in low-$O_2$ furnace environment without presence of surface over-oxides, confirmed by X-ray diffraction and temperature-dependent electrical resistivity measurements. Additional $SiO_2$ layers are deposited above and below the $VO_2$ layer to increase the solar absorption when $VO_2$ is in metallic phase during daytime and to enhance the infrared emissivity when $VO_2$ is in insulating phase at nighttime, revealed by the temperature dependent spectral characterizations. An outdoor vacuum test setup with two optical viewports is custom designed with minimized parasitic heat loads. A contactless infrared camera is used for the sample temperature measurement after careful calibration, while the atmosphere temperature is fitted in-situ at nighttime with a black sample to provide more accurate thermal modeling for validation. The stagnation temperatures of all three $VO_2$ metafilm coatings are measured, where significant temperature rise during the day and temperature drop at nighttime are demonstrated. Energy harvesting performance is theoretically analyzed with heating and cooling powers at various temperature differences.

## 2. Results and Discussion

The concept of an ideal self-adaptive all-day energy-harvesting coating is illustrated in Fig. 1a). During daytime, it is a high-temperature selective absorber with perfect absorption within solar spectrum ($\lambda$ = 0.3~2.5 μm) to maximize solar heating and zero infrared emissivity to minimize the radiative loss. On the other hand, at nighttime it is a sub-ambient selective cooler with blackbody emission to cold outer space (3 K) only within atmospheric transparent window ($\lambda$ = 8~14 μm) and zero infrared absorption elsewhere ($\lambda$ <8 μm and $\lambda$ >14 μm) to eliminate the heating from the atmosphere. With the phase transition temperature $T_c$ = 68°C, thermochromic $VO_2$ is an excellent material to realize temperature-triggered self-switching coatings with its metallic phase ($T_s > T_c$) during daytime and insulating phase ($T_s < T_c$) at nighttime. The temperature rise during daytime and drop at nighttime between the self-adaptive coating ($T_s$) and the ground ($T_w$) could result in self-adaptive daytime solar heating and nighttime radiative cooling, or even direct power generation with solid-state thermoelectric devices.

Figure 1b) illustrates the structural designs and simulated spectral absorptivity($\alpha_\lambda$)/emissivity($\varepsilon_\lambda$) of self-adaptive $VO_2$ metafilm coatings on three different substrate materials, i.e., insulator (Quartz), semiconductor (UDSi) and metal (Al). 300 nm thickness is chosen for the $VO_2$ layer as it is opaque at metallic phase with low infrared emissivity whereas it is semi-transparent in the infrared at insulating phase to allow high phononic thermal emission from underneath $SiO_2$. On top of the 300-nm $VO_2$ layer, 100-nm $SiO_2$ is added as anti-reflection coating in order to achieve high absorption within solar spectrum.[36, 37] Note that Quartz substrate is considered with 500 μm thickness, while UDSi substrate has a thickness of 280 μm based on commercially available wafers. On the other hand, as 200-nm-thick Al is opaque within the solar and infrared spectrum, it is considered semi-infinite in the optical modeling (See Experimental Methods for details), while practically it is deposited on a polished silicon wafer. Finally, the backside of the wafer substrates is coated with highly reflective 200-nm Al film as well for the opaqueness of the sample as well as minimized thermal emission loss. With $VO_2$ in the metallic phase, all three $VO_2$ metafilm coatings exhibit excellent spectral selectivity with high spectral absorptivity up to 0.99 with the solar spectrum and infrared emissivity dropping down to 0.20 in the long wavelength, suggesting potential exceptional solar thermal energy harvesting during daytime. On the other hand, with insulating $VO_2$ phase, the self-adaptive coatings are also spectrally selective with high emissivity up to 0.98 within the atmospheric transparency window ($\lambda = 8\sim14$ μm) and low spectral emissivity down to 0.05 in the short wavelength ($\lambda =2.5\sim8$ μm), indicating potential superior radiative sky cooling performance at nighttime. The oscillating spectra are due to the wave interference from multiple internal reflection within the multilayers. Note that, upon $VO_2$ phase transition, the spectral emissivity within the atmospheric transparency windows could vary by up to 0.8 with these self-adaptive $VO_2$ metafilms on different substrate materials.

These designed $VO_2$ metafilms are successfully fabricated in the process as illustrated in Fig. 2a). High-quality $VO_2$ layer is thermally grown from sputtered 150-nm vanadium films with the cost-effective quartz-tube furnace oxidation method in extremely low-$O_2$ environment (~15 ppm), while $SiO_2$ thin films are coated with plasma enhanced chemical vapor deposition (See Experimental Section for detailed fabrication process). Optical images show the fabricated self-adaptive $VO_2$ metafilm coatings on Quartz, UDSi, and Al substrates all in 1-inch square size. As shown in Fig. 2b), XRD scans confirmed polycrystalline $VO_2$ after furnace oxidation with peaks at 28° (011), 55.5° (211), and 57.5° (022), while temperature-dependent electrical resistivity measurements (Figure 2c) verified the insulator-to-metal phase transition of the $VO_2$ thin films oxidized in low-$O_2$ furnace with change by more than 3 orders of magnitude, which is comparable to the best quality of $VO_2$ grown via MBE method.[38] Figure 2d) shows the measured solar absorptivity and infrared emissivity spectra at 25°C for the insulating phase and at 95°C for the metallic phase of all three fabricated $VO_2$ metafilm coatings, from which excellent spectral selectivity is observed for the self-adaptive coatings as predicted from the modeling (see Figure S1 in the Supplemental Materials for direct comparison to the modeling). When the $VO_2$ is at its insulating phase for nighttime radiative cooling, the $VO_2$/Quartz, $VO_2$/UDSi and $VO_2$/Al coatings exhibit a high in-band total emissivity $\varepsilon_{8-14}$ (i.e., integrated normal emissivity spectrum over blackbody spectrum at 25°C) within the atmospheric transparency window of 0.74, 0.75 and 0.79, respectively. In the short wavelength ($\lambda =2.5\sim8$ μm), $VO_2$/Al and $VO_2$/UDSi samples have very low total emissivity $\varepsilon_{2.5-8}$ of 0.06 and 0.11, while $VO_2$/Quartz has a high emissivity of 0.69 due to much more phonon absorption in the much thicker substrate. On the other hand, all three samples possess high total emissivity $\varepsilon_{14-20}$ around 0.85 for $VO_2$/Quartz and $VO_2$/Al and 0.74 for $VO_2$/UDSi in the long wavelength ($\lambda$

=14~20 μm). When the $VO_2$ is at its metallic phase for daytime solar thermal heating, all three fabricated self-adaptive coatings have the same high total solar absorptivity (i.e., integrated normal absorptivity over air mass 1.5 solar irradiance intensity spectrum) of 0.86 and low total infrared emissivity of around 0.20 ($\lambda$ =2~20 μm). With the temperature dependent FTIR measurement at every 5°C intervals (2°C intervals between 55°C and 75°C, see Fig. S2 in the Supplemental Materials), Figure 2 e) reveals the self-adaptive behavior of the $VO_2$ metafilms on different substrates in terms of in-band emissivity $\varepsilon_{8-14}$ within atmospheric transparent window, where a large variation of 0.58 is observed for $VO_2$/Al sample or 0.55 for $VO_2$/Quartz and $VO_2$/UDSi samples. The shifted temperature dependence of about 10~12°C between heating or cooling is associated with the known IMT thermal hysteresis of $VO_2$. From the emissivity derivative and resistivity derivative with respect to temperature (See Fig. S3 and S4 in the Supplemental Materials), average critical temperature $T_c$ of 67.2°C is obtained, which is consistent with typical $VO_2$ phase transition temperature.[14] Note that, since the plasma frequency of metallic $VO_2$ is located around 1.25 μm wavelength, the optical properties of $VO_2$ and the measured solar absorptivity spectra of the $VO_2$ metafilms have much smaller change upon phase transition in the visible and near-infrared compared to that in the mid-infrared (See Fig. S5 in the Supplemental Materials).

To demonstrate the self-adaptive all-day energy harvesting with the tunable $VO_2$ metafilm coatings, outdoor tests are carried out for measuring the stagnation temperatures (i.e., sample temperature with zero net heat transfer) at both daytime and nighttime in a home-built vacuum setup with two viewports (KBr and ZnSe), as pictured in Fig. 3a), where convective effect is eliminated. A pyranometer is set aside in ambient to measure the solar irradiance outside the vacuum setup during daytime. Moreover, to minimize the parasitic heat transfer, samples sit on thin wood sticks and are placed as close as to the viewports. As depicted in Fig. 3b), an infrared camera (See Fig. S6 in the Supplemental Materials for validation of temperature measurement) is used for measuring the sample temperature in a noncontact fashion without attaching thermal sensors on the sample surfaces. Instead of taking it as same as the ambient temperature, effective atmospheric temperature ($T_{atm}$) is fitted in-situ with a black sample (0.99 solar absorptivity and infrared emissivity) under the ZnSe viewport along with the test sample under the KBr viewport. KBr viewport is chosen for $VO_2$ metafilms due to high transmittance in solar spectrum while ZnSe viewport with high transmittance in 8-14 μm spectrum is chosen for the black sample. As the black sample interacts radiatively with both the atmosphere and the outer space, effective atmospheric temperature at nighttime can be found based on the heat transfer model using the measured black sample temperature ($T_{s,b}$) (See Experimental Section for detailed energy balance equation and fitting method). As illustrated in Figure 3c), major heat transfer at nighttime includes thermal emission from the sample ($Q_{emiss}$) and absorbed thermal radiation from atmosphere ($Q_{atm}$), while parasitic heat transfer includes radiative heat exchange between the VO$_2$ coating and vacuum wall ($Q_{s-w}$) and between sample backside and the vacuum wall ($Q_b$) as well as conductive heat transfer through wood sticks ($Q_{wood}$) and through conduction by air molecules ($Q_{air}$). Fitting nighttime atmosphere temperature with in-situ experimental data is crucial for accurate quantification of $Q_{atm}$ and heat transfer modeling. Using the fitted $T_{atm}$ from the black sample, the stagnation temperatures of VO$_2$ metafilm samples are calculated for validating the measured values from outdoor vacuum tests.

The measured nighttime stagnation temperature drop $\Delta T$ (i.e., $T_s - T_w$) of black sample and VO$_2$ metafilms under KBr viewport is shown in Figure 3d) in excellent agreement with the modeling at measured vacuum pressures. Black sample with almost-unity emissivity achieved maximum of −19°C drop from the wall temperature ($T_w \sim$ 30°C) at 1 Pa vacuum pressure, whereas VO$_2$/Quartz, VO$_2$/UDSi and VO$_2$/Al sample reached −12°C, −15°C, and −17°C drop, respectively. Thanks to the lowest short-wavelength emissivity ($\varepsilon_{2.5-8}$) and highest in-band emissivity ($\varepsilon_{8-14}$), the VO$_2$/Al sample achieved largest temperature drop among all three VO$_2$ metafilm samples. However, all VO$_2$ metafilms do not exceed temperature drop observed by the black emitter, not only because their in-band emissivity ($\varepsilon_{8-14}$ = 0.74~0.79) is lower than the black emitter (0.99) but also due to the air molecule conduction associated with the vacuum pressures (1.0~3.4 Pa) during the nighttime tests. Note that the nonnegligible heat gain via air conduction normalized to emitted thermal radiation from the sample ($Q_{air}/Q_{emiss}$) is respectively 5%, 7%, 12%, and 8% for black sample, VO$_2$/Quartz, VO$_2$/UDSi, and VO$_2$/Al, while other parasitic heat gains (i.e., $Q_{s-w}$, $Q_b$, $Q_{wood}$) are negligible with less than 1% of $Q_{emiss}$ (See Figure S7 in the Supplemental Material). The heat gain is dominated by the radiative heat flux absorbed from the atmosphere $Q_{atm}$ with 85~90% of $Q_{emiss}$ for VO$_2$ metafilms (See Figure S8 in the Supplemental Material). As black sample interacts more with the atmosphere, its dominance is the largest among others with ~93%, regardless of its highest radiative heat flux from the sample ($Q_{emiss}$~260 W/m$^2$). It shows that the atmosphere temperature $T_{atm}$ becomes crucial during nighttime because of nearly comparable heat flux from the atmosphere ($Q_{atm}$~240 W/m$^2$ for black). Hence, atmosphere temperature $T_{atm}$ needs to be fitted such that emission from the atmosphere is accurately evaluated. With 0 Pa vacuum pressure where there is no air conduction, the fabricated VO$_2$ samples have a potential to reach maximum temperature drop of −25°C, which is 5°C lower than that of black emitter, indicating the significance of spectral selectivity. An ideal selective emitter can achieve −40°C to −30°C temperature drop depending on the weather conditions with negligible air conduction inside the vacuum chamber.

During daytime tests, samples inside the vacuum chamber are heated up by solar irradiation ($Q_{sun}$) passing through the viewports, as depicted in Figure 4a). Around noon time, the black sample ($\alpha_{sol}=\varepsilon_{IR}=0.99$) under the KBr viewport reaches a maximum temperature rise $\Delta T$ (i.e., $T_s - T_w$) of 62°C above the wall temperature, as shown in Figure 4b). On the other hand, all three VO$_2$ metafilm samples under KBr viewport achieve much higher temperature rises of 169°C, 168°C and 156°C respectively on Quartz, UDSi and Al substrates, thanks to their self-adaptive spectral selectivity ($\alpha_{sol}=0.86$ and $\varepsilon_{IR}\sim 0.2$) after VO$_2$ changes to its metallic phase above the critical temperature ($T_c = 68°C$) upon solar heating. Note that VO$_2$/Al sample achieves relatively smaller temperature rise because of its slightly higher infrared emissivity than other two VO$_2$ metafilm samples. Thermal modeling is performed to validate the daytime stagnation temperature measurements where excellent agreement is observed. In particular, the atmosphere temperature ($T_{atm}$) during daytime is estimated based on the measured vacuum wall temperature ($T_w$) assuming that their difference ($T_w-T_{atm}$) does not change between nighttime and daytime. In addition, a correction factor $c_1$ is considered for the attenuation when the sunlight passes through the viewports with oblique incidence at different time of the day due to change in solar angles. During the VO$_2$ metafilm test under the KBr viewport, a black sample is measured under ZnSe viewport for fitting time-dependent $c_1$ factors simultaneously (See Experimental Section for $c_1$ factor fitting). At the peak solar irradiance, the $c_1$ factor is 0.69 for the KBr viewport and 0.54 for the ZnSe viewport. A detailed heat transfer analysis reveals that, radiative losses from the backside ($Q_b$) and from the top surface to the vacuum wall ($Q_{s-w}$) are about 7% and 6% respectively to the absorbed solar energy ($Q_{sun}$) for VO$_2$ metafilm samples (See Figure S9 in the Supplemental Material). Parasitic conductive heat losses via wood sticks ($Q_{s-w}$) and air molecules under 0.03 Pa during the daytime tests are negligible as both are less than 1% of $Q_{sun}$. Further lowering the vacuum pressure can only improve the temperature rise by up to 1°C according to the modeling (See Figure S10 in the Supplemental Material). The emitted heat flux from the sample ($Q_{emiss}$) is comparable to the absorbed solar radiation ($Q_{sun}$), resulting in high stagnation temperature for VO$_2$ metafilms ($Q_{emiss}/Q_{sun}\sim 1.0$) (See Figure S11 in the Supplemental Material). While the heat gain from the atmosphere is only 15% in value compared to absorbed solar radiation ($Q_{atm}/Q_{sun}\sim 0.15$). On the other hand, black emitter absorbs more solar energy, but at the same time, its emitted energy is larger due to broadband emissivity of $\varepsilon_{IR} = 0.99$. The emitted thermal radiation exceeds that of received solar irradiance more than 1.5 times ($Q_{emiss}/Q_{sun}>1.5$) for the black sample. Due to the interaction with atmosphere, black sample also absorbs heat from the atmosphere, whose heat flux is more than half of absorbed solar radiation ($Q_{atm}/Q_{sun}>0.5$). For an ideal selective solar absorber, stagnation temperature can reach ~350°C above the wall temperature in vacuum under the same solar irradiance and weather conditions.

To fully understand the energy-harvesting performance, Figure 5a) calculates the theoretical cooling power ($q_{net}$) at nighttime ($T_w = 30$°C and $T_{atm} = 18$°C) as a function of temperature drop $\Delta T = T_s - T_w$ for all three self-adaptive $VO_2$ metafilms, the black emitter, and ideal selective emitter under KBr viewport in vacuum with negligible parasitic losses of $Q_{air}$ and $Q_{wood}$. At equilibrium temperature (i.e., $T_s = T_w$), the $VO_2$ metafilm coatings could achieve a maximum cooling power of ~60 W/m² comparable to the ideal selective emitter, while the broadband black emitter could dissipate a maximal heat flux of 86 W/m². The slightly more cooling power that the black emitter exhibits is due to the additional heat dissipation outside the atmospheric transparency window to the atmosphere whose temperature is 12°C lower. As a result, the black emitter outperforms ideal selective emitter with the temperature drop down $\Delta T$ from 0 to −10°C, or the $VO_2$ metafilms down to −20°C around which the stagnation temperature with zero cooling power exists. Overall, the black sample exhibits a sharp slope due to its broadband near-unity emissivity, whereas all three $VO_2$ metafilms behave very similarly with shallow slopes affected by much lower short-wavelength emissivity ($\varepsilon_{2.5\sim 8}$). Note that $VO_2$/UDSi shows slightly lower stagnation temperature than the other two $VO_2$ metafilms because of its lower long-wavelength emissivity ($\varepsilon_{2.5\sim 8}$). Under the same conditions, the ideal selective emitter can reach the lowest stagnation temperature of −4°C (or largest temperature drop $\Delta T = -34$°C) because of shallower slope from high spectral selectivity with both zero short- and long-wavelength emissivity outside of atmospheric transparency window.

During daytime at peak solar irradiance ($q_{sun} = 1017$ W/m², $c_1 = 0.69$, $T_w = 50$°C, $T_{atm} = 38$°C), heating power is calculated as a function of temperature rise $\Delta T$ as shown in Figure 5b). As our fabricated $VO_2$ thin film turns into complete metallic phase at 80°C, the lowest temperature rise is kept at 30°C in order to ensure the VO$_2$ metafilms adapted as selective solar absorbers. With 30°C temperature rise above the wall temperature, three VO$_2$ metafilms reach a heating power around 400 W/m$^2$, which is higher by 140 W/m$^2$ than the black sample and lower by only 60 W/m$^2$ than the ideal selective solar absorber. As the temperature increases, while the heating power starts to decrease, all the VO$_2$ metafilms significantly outperform the black sample, thanks to the excellent spectral selectivity ($\alpha_{sol}$=0.86 and $\varepsilon_{IR}$~0.2) enabled by the metallic phase of VO$_2$. The maximal temperature rise at stagnation temperature where there is no heating power (i.e., $q_{net}$ = 0) is 170°C for VO$_2$/UDSi and VO$_2$/Quartz samples, 156°C for VO$_2$/Al sample, and only 61°C for the black sample, which are consistent with the experimental observation from the outdoor vacuum tests. The VO$_2$/Al sample slightly underperforms than the VO$_2$/Quartz and VO$_2$/UDSi because of slightly higher infrared emission loss. Note that, only 57% of solar irradiation reaches the sample surface after passing through the KBr viewport due to the solar tilting angle ($c_1$ = 0.69) and non-unity solar transmittance ($\tau_v$~0.83, see Figure S12 for normal transmittance spectrum). It is expected that heating power up to 680 W/m$^2$ at $\Delta T$ = 30°C and maximal temperature rise up to 520 °C at $q_{net}$ = 0 could be achieved with the VO$_2$ metafilm coatings if they are orientated normally to the Sun by solar tracking with more transparent viewport.

## 3. Conclusion

This work has demonstrated self-adaptive energy harvesting with tunable vanadium dioxide (VO$_2$) metafilms fabricated on different substrate materials with outdoor vacuum tests. During daytime, a large temperature rise up to 169 K is achieved due to the high solar absorptivity of 0.86 and low emissivity of ~0.2 of the metafilms with VO$_2$ in its metallic phase. At nighttime, 17 K temperature drop is obtained with VO$_2$ in its insulating phase, where it possesses a high in-band emissivity $\varepsilon_{8-14}$ of 0.79 and a short-wavelength emissivity $\varepsilon_{2.5-8}$ as low as 0.06. Accurate heat transfer analysis predicts a peak heating power of 400 W/m$^2$ at 80°C sample temperature with current experimental setup, which could be further improved up to 680 W/m$^2$ with solar tracking and more transparent viewport. The nighttime cooling power could reach 60 W/m$^2$ when the metafilms are in equilibrium with ambient at 30°C. It is expected that these tunable metafilms could continuously generate power all day long with daytime solar heating and nighttime radiative cooling effects when paired with TEGs or other solid-state power generators as novel sustainable and renewable power sources.

## 4 Experimental Section

*Design of VO$_2$ metafilms*: The complex refractive index of VO$_2$ used in the expected spectra is obtained by fitting from synthesized 300-nm-thick VO$_2$ on undoped- and doped-silicon wafers in the low-O$_2$ quartz tube furnace (See Fig. S13 in the Supplemental Material). The refractive index *n* and extinction coefficient *κ* of VO$_2$ are extracted at a given wavelength by minimizing the absolute difference of absorptance between the measured and theoretical spectra as the objective function with Bayesian optimization. The complex refractive indices of SiO$_2$ layer, Quartz, undoped Si, and Al are taken from Palik's handbook.[39] The absorptivity and emissivity spectra of the VO$_2$ metafilm on thick substrates are calculated with transfer matrix method that considers wave interference in multilayers coupled with the ray-tracing approach that takes into account the internal transmittance within the thick substrates.[40]

*Fabrication of VO$_2$ metafilms*: it includes four steps with aluminum deposition, SiO$_2$ deposition, vanadium deposition and low-O$_2$ oxidation. For VO$_2$/Al, 200-nm-thick aluminum is sputtered (Lesker PVD75 sputterer) on both sides of double-side polished silicon wafer (2-inch in diameter, Universitywafer Inc.) at a rate of 1.0 Å/s with DC power of 160 W in vacuum pressure of $5 \times 10^{-6}$ Torr and process pressure of 4 mTorr with Ar gas flow. Here, the aluminum covered silicon wafer is denoted as Al substrate. Similarly, 200-nm-thick aluminum is sputtered on back surfaces of 1-inch square Quartz (z-cut, MSE Supplies) and 2-inch diameter undoped silicon (UDSi, $\rho$>10000 Ω·cm, Universitywafer Inc.) double-side polished wafers. SiO$_2$ of 2-μm thick is deposited via plasma enhanced chemical deposition (PECVD, Oxford) on front sides of UDSi wafer for VO$_2$/UDSi and VO$_2$/Al sample. Note that 100 nm SiO$_2$ is deposited as anti-reflection coating, once 300 nm VO$_2$ is formed on top of 2μm-thick SiO$_2$ and Quartz substrate. Vanadium (purity > 99.9% sputtering target, Kurt J. Lesker) is sputtered onto three different structures: (i) Quartz / 200-nm-thick aluminum, (ii) 2-μm-thick SiO$_2$ / UDSi / 200nm-thick aluminum and (iii) 2-μm-thick SiO$_2$ / Al at a rate of 0.9 Å/s with DC power of 125 W in vacuum pressure of $5 \times 10^{-6}$ Torr and process pressure of 4 mTorr with Ar gas flow (Lesker PVD75 sputterer). 10 mins is given to reach stable deposition rate before opening the shutter for 27 mins to deposit about 150 nm vanadium thin film onto the substrates. An adequate time of about 30 mins is also given for the samples to sufficiently cool down to prevent any oxidation after venting the vacuum chamber. The sputtered 150-nm vanadium is optimally oxidized at 500°C in controlled low-oxygen furnace for 12 hours. Nitrogen gas (purity > 99.99%) is purged at a flowrate of 1.0 liter-per-minute (LPM) throughout the oxidation process which maintains the residual $O_2$ content at ~15 ppm inside the furnace (see Figure S14 in the Supplemental Material).

*Material characterization*: The thickness of deposited vanadium and oxidized vanadium film is measured with profilometer (Bruker DektakXT, See Figure S15 for thickness measurement). Temperature-dependent infrared reflectance of $VO_2$ samples is measured with an FTIR spectrometer (Nicolet iS50, Thermo Fisher Scientific) at near normal incident angle of 10º with reflectance accessory (Seagull, Harrick Scientific) from 2 μm to 20 μm in wavelength with 4 cm$^{-1}$ resolution and each spectrum averaged over 64 scans (See Figure S16 for reflectance measurement). Aluminum mirror is used as the reference for reflection measurement. Similarly, temperature-dependent solar reflectivity is measured with a tunable light source (TLS) and a PTFE integrating sphere equipped with home-made temperature stage (See Figure S17 in the Supplemental Material). Si detector is used for visible range (0.35 ~ 0.9 μm) and InGaAs detector is used for near-infrared range (0.95 ~ 1.65 μm). Temperature-dependent electrical resistivity is measured with a home-made four-probe station (See Figure S18 for the setup and validation) along with a source meter (Keithley 2401). The resistivity is calculated from the slope of V-I curve (*S*) and thickness (*t*) as $\rho = C\pi S t/ln2$, where *C* is the geometrical factor. Sample temperature is varied with a custom built temperature stage made of a thermoelectric heater and a PID temperature controller along with K-type thermocouples. Heating-cooling curves are measured from 25°C (insulating phase) to 95°C (metallic phase). Steady state condition is achieved after the sample temperature reaches setpoint for 5 mins with fluctuation less than ±1°C. Lastly, grazing incidence X-ray diffraction (GIXRD, Rigaku SmartLab) is performed with Cu K$\alpha$ source. The 2$\theta$ scans are conducted with angle of incidence $\omega$ = 0.5° to probe the surface of the thin film. The obtained diffraction patterns were compared with database (HighScore Plus).

*Stagnation temperature measurement*: Stagnation temperature is measured with infrared camera (FLIR E8-XT, Teledyne FLIR) every 15 minutes. Infrared camera is held approximately 18 inches above the viewport to minimize blockage of sunlight and its temperature readings have been validated with reference sample (black sample and heavily doped silicon, See Figure S6 in the Supplemental Material) sitting under ZnSe and KBr viewports in the lab. The measured temperature is analyzed with the software (FLIR ignite) by defining the emissivity (7.5 – 13 μm) of the sample and reflected apparent temperature. Aluminum is used to determine the reflected apparent temperature, which takes into account the effect of background noise.

*Energy balance equation*: The steady-state heat transfer model is used to predict the stagnation temperature of the sample under outdoor vacuum test when net heat transfer is zero ($Q_{\text{net}} = 0$):

$$Q_{\text{sun}} + Q_{\text{atm}}(T_s, T_{\text{atm}}) - Q_{\text{emiss}}(T_s) - Q_{\text{para}}(T_s, T_w) = Q_{\text{net}} \qquad (1)$$

Absorbed solar energy through the viewport is calculated as $Q_{\text{sun}} = c_1 \tau_v \alpha_{\text{sol}} A_s q_{\text{sun}}$ where $q_{\text{sun}}$ is the solar irradiance measured by the ambient pyrometer at daytime. Total solar absorptivity is $\alpha_{\text{sol}} = \frac{1}{G_{\text{AM1.5}}} \int \alpha_{s,\lambda}(T_s) G_{\text{AM1.5},\lambda} d\lambda$ integrated over spectral solar irradiance ($G_{\text{AM1.5},\lambda}$)

Radiative heating by atmosphere is calculated as $Q_{atm}(T_s, T_{\text{atm}}) = 2\pi A_s c_2 \int_0^{\pi/2} \int \tau_{v,\lambda} \varepsilon_{s,\lambda}(T_s) \varepsilon_{\text{atm},\lambda} I_{\text{BB},\lambda}(T_{\text{atm}}) \cos\theta \sin\theta \, d\lambda d\theta$, where $\varepsilon_{s,\lambda} = 1 - R_{s,\lambda}$ is the spectral infrared emissivity of the sample calculated from measured $R_{s,\lambda}$ at near-normal direction, $\varepsilon_{\text{atm},\lambda} = 1 - \tau_{\text{atm},\lambda}^{(1/\cos\theta)}$ is the spectral emissivity of the atmosphere with atmospheric transmittance $\tau_{\text{atm},\lambda}$ extracted from Planetary Spectrum Generator (psg.gsfc.nasa.gov) at the test location (Tempe, AZ, USA, 33.4°N, 111.9°W) for the given date, $I_{\text{BB},\lambda}$ is the spectral blackbody emissivity at the atmospheric temperature $T_{\text{atm}}$ fitted with the in-situ measurement data from the black sample under ZnSe viewport, $\tau_{v,\lambda}$ is the normal spectral transmittance of the viewport measured by the FTIR spectrometer in the infrared, and $c_2$ accounts for the ratio between hemispherical and normal transmittance of the viewport from ray-tracing calculation (0.92 for both KBr and ZnSe viewports).

Sample thermal emission through the viewport is calculated by $Q_{emiss}(T_s) = A_s \pi c_2 \int F_{s-v} \tau_{v,\lambda} \varepsilon_{s,\lambda}(T_s) I_{\text{BB},\lambda}(T_s) d\lambda$, where $F_{s-v}$ is the view factor from the sample to the viewport calculated based on the sample size, viewport size, and the distance (i.e., 8 mm for ZnSe and 5 mm for KBr).[41]

Parasitic heat loss/gain ($Q_{\text{para}}$) includes four parts: radiative parasitic heat transfer from the sample front surface to the vacuum wall as $Q_{s-w} = \frac{\sigma(T_s^4 - T_w^4)}{\frac{1-\varepsilon_s}{\varepsilon_s A_s} + \frac{1}{A_s F_{s-w}} + \frac{1-\varepsilon_w}{\varepsilon_w A_w}}$, radiative parasitic heat transfer from sample back surface to the vacuum walls as $Q_b = \varepsilon_b A_s \sigma(T_s^4 - T_w^4)$, conductive parasitic heat transfer via air molecules between the sample and viewport as $Q_{air} = \frac{k_{\text{eff}} A_s}{d}(T_s - T_w)$, and conductive parasitic heat transfer via the wood sticks as $Q_{\text{wood}} = \left(\frac{kA}{L}\right)_{\text{wood}} (T_s - T_w)$. Here, sample-to-wall view factor is $F_{\text{s-w}} = 1 - F_{\text{s-v}}$, $\varepsilon_w$ is the total hemispherical emissivity of aluminum tape attached to the wall, $\varepsilon_b$ is the total hemispherical emissivity of aluminum coated on the backside of the sample, $T_w$ is the wall temperature measured by the K-type thermocouple, $A_s$ and $A_w$ are the surface area of the sample and the wall. In addition, thermal conductivity of wood is taken as $k_{wood} = 0.1$ W/m·K and that of air is calculated based on the measured vacuum pressure.[42]

*Atmosphere temperature fitting*: The atmosphere temperature $T_{atm}$ is fitted from measured $T_{s,ZnSe}$ (black sample) on ZnSe viewport side during nighttime by using Eq. (1). With fitted atmosphere temperature, stagnation temperature of VO$_2$ metafilms ($T_{s,KBr}$) during nighttime is determined by solving the energy balance equation (Eq. (1)) applied to KBr viewport side. Similarly, during daytime, with assumption of difference between atmosphere temperature and wall temperature ($T_{atm} - T_w$) stay the same from nighttime, stagnation temperature is solved.

*$c_1$ factor fitting:* During daytime, correction factor $c_1$ is necessary to solve the stagnation temperature. First, the $c_1$ factor for both viewports is determined with the black sample as $c_1 = Q_{sun,black}/(\tau_v A_s q_{sun})$, where $Q_{sun,black}$ is absorbed solar radiation by the black sample calculated from the heat transfer model using the measured stagnation temperature, $\tau_v$ is the solar transmittance of the viewport at normal incidence, $A_s$ is the sample surface area, and $q_{sun}$ is the measured solar irradiance by the pyrometer placed outside the vacuum setup. Then, time-dependent $c_1$ ratio $r(t) = c_{1,KBr,black}(t)/c_{1,ZnSe,black}(t)$ is obtained from black sample under both KBr and ZnSe viewport. The ratio is used to obtain time-dependent $c_{1,KBr}(t) = r(t)c_{1,ZnSe}(t)$ when VO$_2$ metafilm sits under KBr viewport and where $c_{1,ZnSe}(t)$ is obtained from black sample under ZnSe viewport.

**Supporting Information**
Supporting Information is available from the Wiley Online Library or from the author.

**Acknowledgements**
This work was supported by the National Science Foundation (NSF) under Grant No. CBET-2212342. We would like to thank ASU NanoFab and Goldwater Center for use of their nanofabrication and characterization facilities.

**Conflict of Interests**
The authors do not declare conflict of interest.

**Author Contributions**

K.A. fabricated the samples, characterized optical and electrical properties, measured the stagnation temperatures, prepared the figures, and drafted the manuscript. V.K.R. oxidized the samples and performed XRD scans. L.W. conceived the idea, supervised the work, secured the funding, and revised the manuscript. All authors have reviewed and approved the final version of the manuscript.

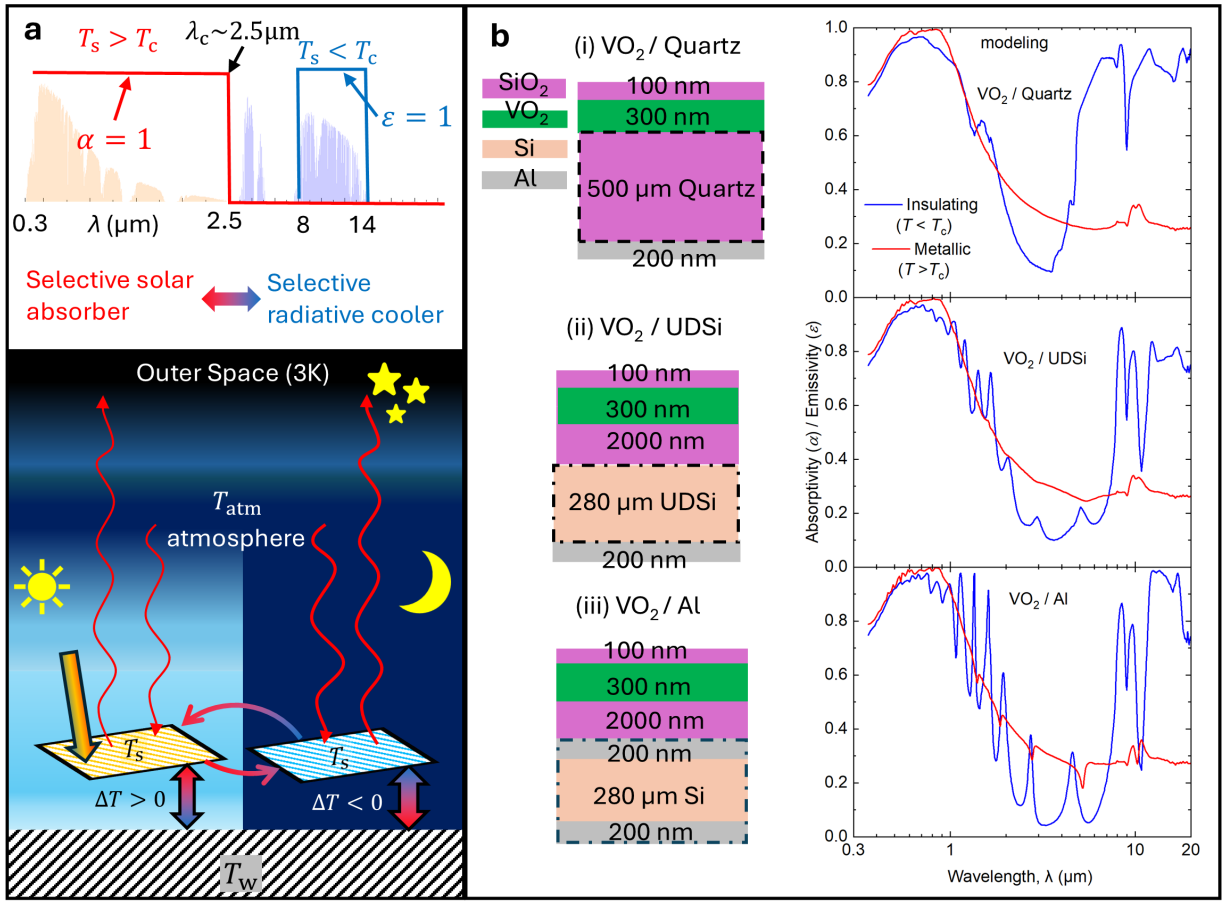

**Figure 1.** a) Schematic of daytime solar heating and nighttime radiative cooling with self-adaptive energy-harvesting coatings. b) Designs of $VO_2$ metafilm self-adaptive coatings on insulator (Quartz), semiconductor (undoped-silicon wafer, UDSi), and metal (Al) coated substrates.

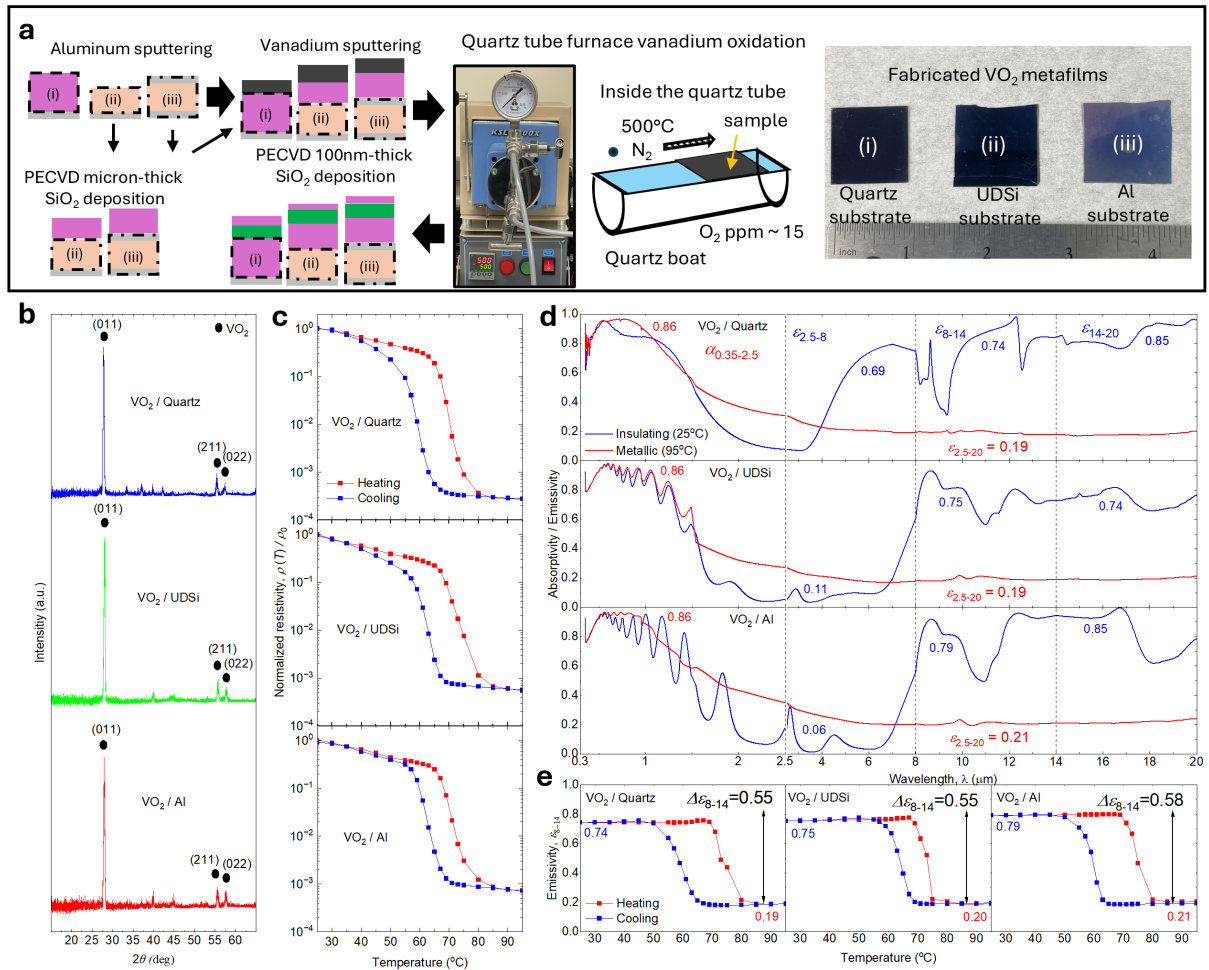

**Figure 2.** a) Flow of fabrication process and pictures of fabricated $VO_2$ metafilm self-adaptive coatings on different substrates, b) XRD scans, c) heating-cooling curves of normalized electrical resistivity d) measured absorptivity and emissivity spectra, and e) heating-cooling curves of in-band emissivity ($\varepsilon_{8-14}$) of $VO_2$ metafilms on Quartz, UDSi, and Al substrates.

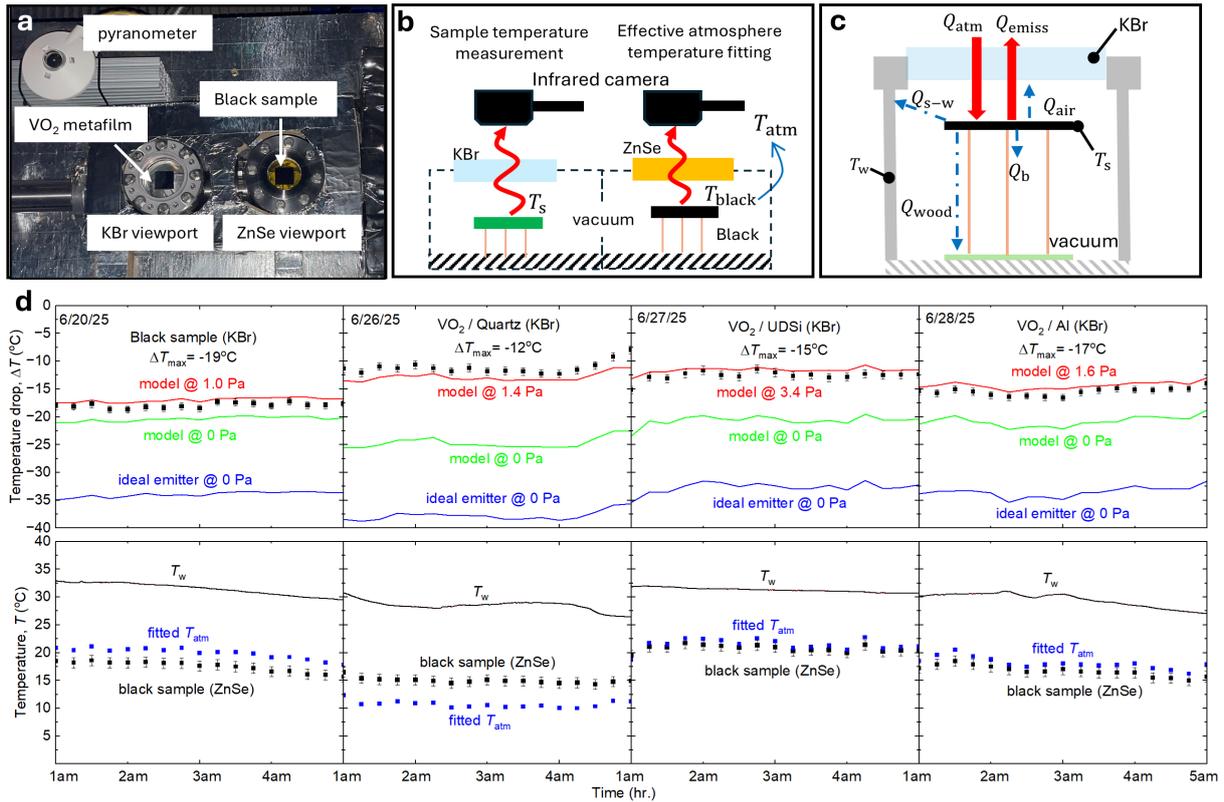

**Figure 3.** a) Picture of setup (view from top). b) Schematic of effective atmosphere temperature fitting and sample temperatrue measurement. c) Schematic of heat transfer model during nighttime. d) Measured and predicted stagnation temperature drop ($\Delta T = T_s - T_w$) (top column) and measured wall temperature ($T_w$) and fitted atmosphere temperature ($T_{atm}$) obtained from stagnation temperaure of black sample (bottom column) and of $VO_2$ metafilm self-adaptive coatings on Quartz, UDSi, and Al substrates.

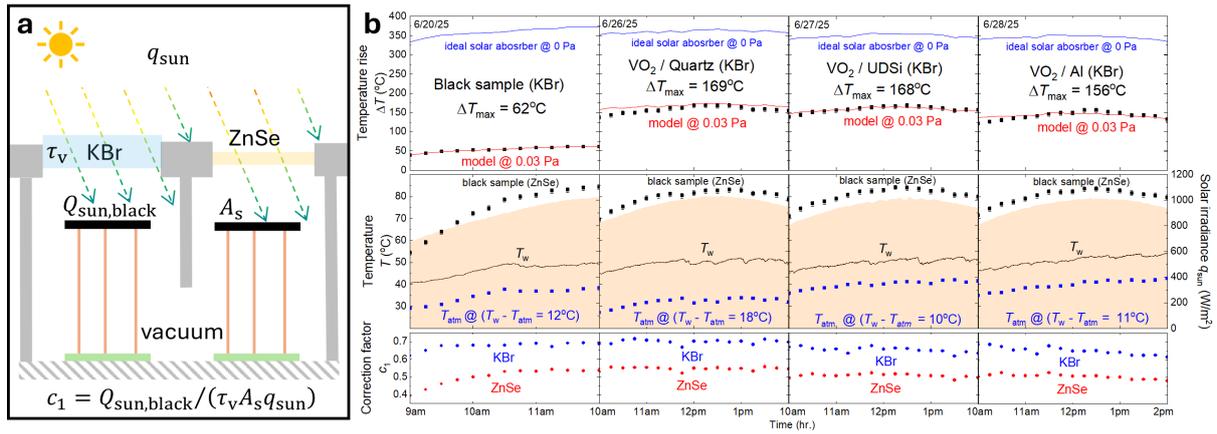

**Figure 4.** a) Schematic of effect on the oblique incidence. b) Measured and predicted stagnation temperature rise ($\Delta T = T_s - T_w$) (top column) and measured wall temperature ($T_w$) and atmosphere temperature ($T_{atm}$) obtained from nighttime stagnation temperaure (bottom column) of $VO_2$ metafilm self-adaptive coatings on Quartz, UDSi, and Al substrates.

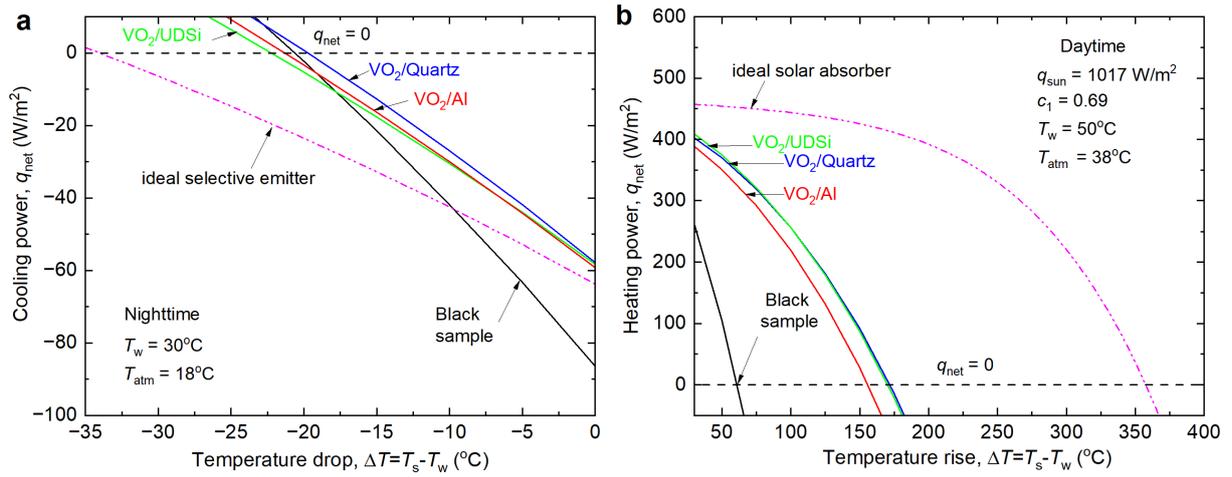

**Figure 5.** a) Theoretical cooling power at nighttime when $VO_2$ is at its insulating phase (wall temperature $T_w = 30°C$, $T_{atm} = 18°C$). b) Theoretical heating power at daytime when $VO_2$ is at its metallic phase ($q_{sun} = 1017$ W/m$^2$, $c_1 = 0.69$, $T_w = 50°C$, $T_{atm} = 38°C$).

21